\documentclass[english]{revtex4}
\usepackage[T1]{fontenc}
\usepackage[latin9]{inputenc}
\usepackage{amsmath}
\usepackage{cancel}
\usepackage{slashed}
\usepackage{graphicx}

\makeatletter

\providecommand{\tabularnewline}{\\}

\@ifundefined{textcolor}{}
{%
 \definecolor{BLACK}{gray}{0}
 \definecolor{WHITE}{gray}{1}
 \definecolor{RED}{rgb}{1,0,0}
 \definecolor{GREEN}{rgb}{0,1,0}
 \definecolor{BLUE}{rgb}{0,0,1}
 \definecolor{CYAN}{cmyk}{1,0,0,0}
 \definecolor{MAGENTA}{cmyk}{0,1,0,0}
 \definecolor{YELLOW}{cmyk}{0,0,1,0}
}

\makeatother

\usepackage{babel}
\begin{document}
\title{Dynamic Structure of Hadrons in ChPT}
\author{A. Aleksejevs and S. Barkanova}
\begin{abstract}
The Chiral Perturbation Theory (ChPT) has been very successful in describing low-energy hadronic properties in the non-perturbative regime of Quantum Chromodynamics. The results of ChPT, many of which are currently under active experimental investigation, provide stringent predictions of many fundamental properties of hadrons, including quantities such as electromagnetic polarizabilities. 
The paper outlines our semi-automated calculations in ChPT, the corresponding results for the electric and magnetic polarizabilities of the proton and our predictions for Compton differential cross sections.
\end{abstract}
\maketitle
\section{Formalism and Results}
\begin{figure}
\begin{centering}
\begin{tabular}{cc}
\includegraphics[scale=0.4]{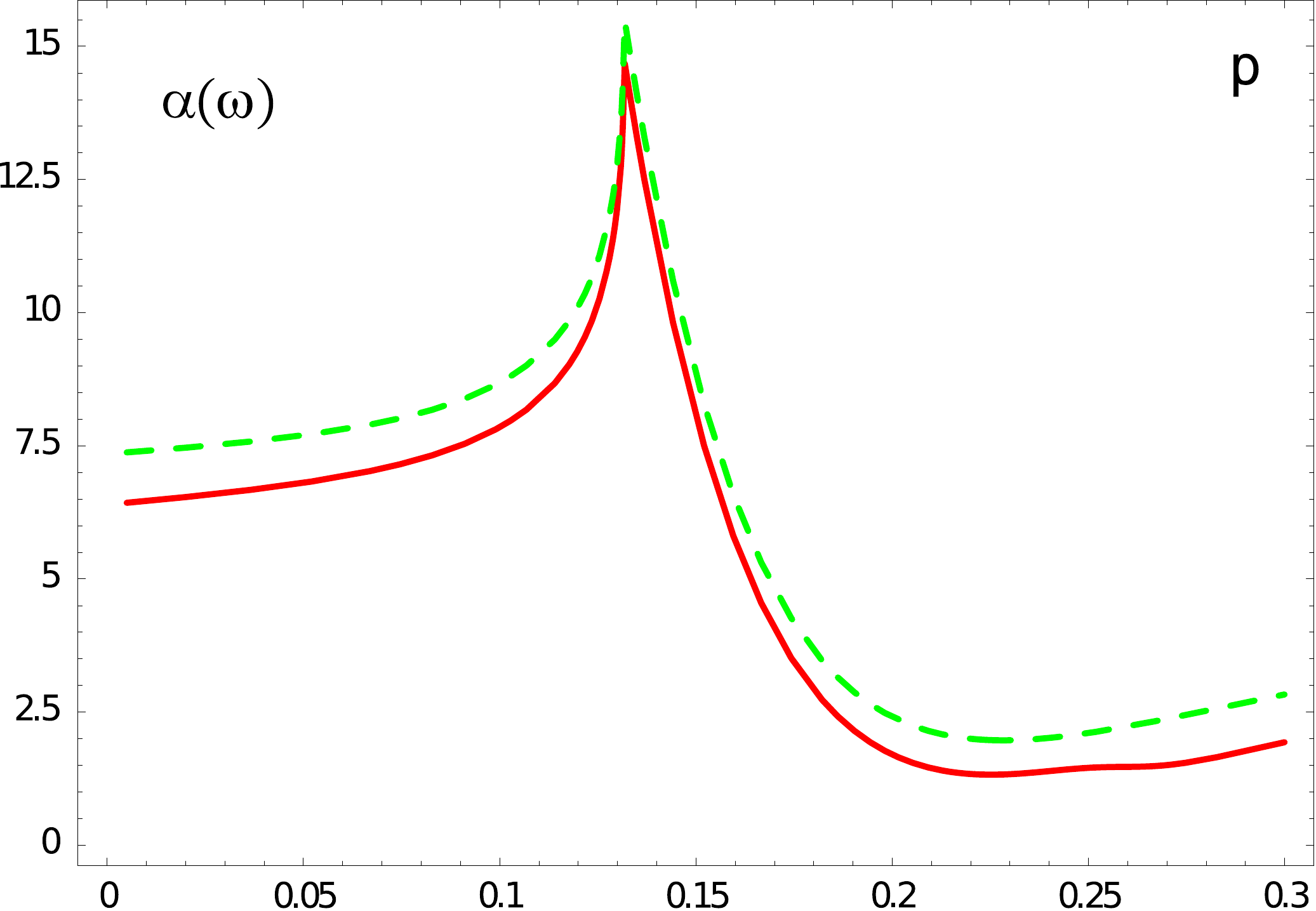} & \includegraphics[scale=0.4]{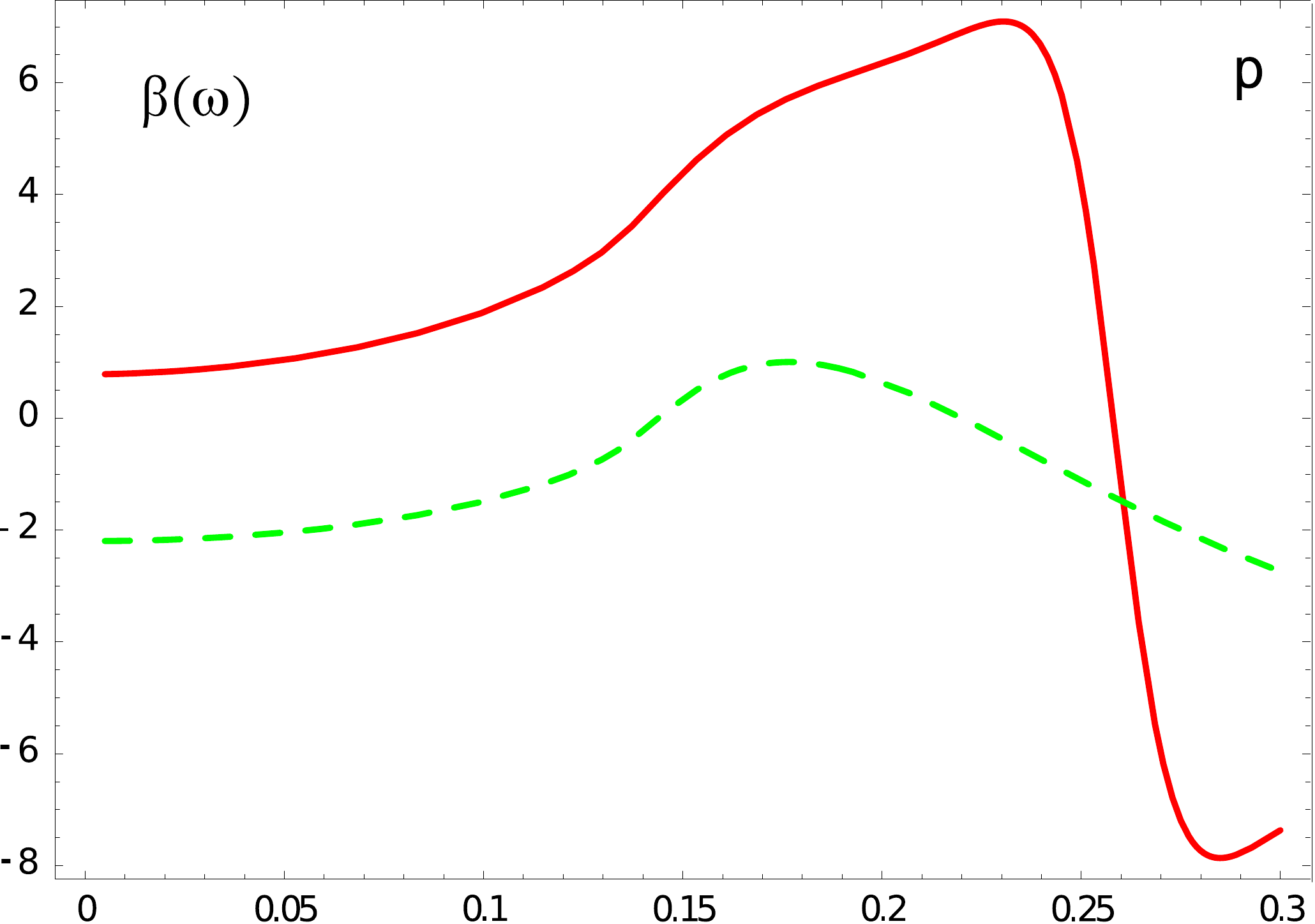}\tabularnewline
 & \tabularnewline
\end{tabular}
\par\end{centering}
\centering{}\caption{Dependencies of the proton electric and magnetic polarizabilities
(in $10^{-4}\,(fm^{3})$) on photon energy $\omega$ (GeV) in c.m.s. The green-dashed curves correspond to
the $\mathcal{O}\left(p^{3}\right)-\pi N$ loops contribution and
WZW-anomaly. The solid-red curves include all the previous contributions
plus the $\Delta$-pole resonance.}
\label{ff1}
\end{figure}

In general, polarizabilities are related to the deformability and stiffness
of hadrons and can be experimentally accessed through Compton
scattering. For the proton and neutron, electric ($\alpha$) polarizabilities
are approximately the same, while magnetic ($\beta$) polarizabilities are
different but both positive, which points to
the paramagnetic nature of the nucleon. Although small (on order of $10^{-4}\, fm^{3}$), electric and magnetic polarizabilities were measured by several experimental groups. We can relate an amplitude to the set of Compton structure functions $R_{i}$ \cite{Babusci} in the following way: 
\begin{align}
\frac{1}{8\pi W}M(\gamma B\rightarrow\gamma^{\prime}B)= & R_{1}({\bf\epsilon}^{\prime*}\cdot{\bf\epsilon})+R_{2}({\bf s}{}^{\prime*}\cdot{\bf s})+iR_{3}{\bf\sigma}\cdot({\bf\epsilon}{}^{\prime*}\times{\bf\epsilon})+iR_{4}l{\bf\sigma}\cdot({\bf s}{}^{\prime*}\times{\bf s}) \label{eq:1} \\
 & + iR_{5}(({\bf\sigma}\cdot\hat{{\bf k}})({\bf s^{\prime*}}\cdot{\bf\epsilon})-({\bf\sigma}\cdot\hat{{\bf k}}^{\prime})({\bf s}\cdot{\bf\epsilon^{\prime*}}))+iR_{6}(({\bf\sigma}\cdot\hat{{\bf k}}^{\prime})({\bf {\bf s^{\prime*}}}\cdot{\bf \epsilon})-({\bf\sigma}\cdot\hat{{\bf k}})({\bf s}\cdot{\bf \epsilon^{\prime*}})).\nonumber
\end{align}
Here, $W=\omega+\sqrt{\omega^{2}+m_{B}^{2}}$ is the center of mass energy and $\omega$ is the energy of the incoming photon. Unit magnetic
vector (${\bf s}=(\hat{{\bf k}}\times{\bf\epsilon}$)), polarization
vector ($\boldsymbol{\epsilon}$) and unit momentum of the photon
$\big( \hat{{\bf k}}={\bf k}/k\big)$ are denoted by
the prime for the case of the outgoing photon. In our case we have
computed the Compton scattering amplitude using CHM \cite{CHM} in the basis of Dirac chains:
\begin{align}
\frac{1}{8\pi W}M(\gamma B\rightarrow\gamma^{\prime}B)= & f_{1}(\epsilon^{\prime*}k)\left[\overline{u}\left(p^{\prime}\right) \slashed{\epsilon}u\left(p\right)\right]+f_{2}(\epsilon k^{\prime})\left[\overline{u}\left(p^{\prime}\right) \slashed{\epsilon}^{\prime*}u\left(p\right)\right]+f_{3}(\epsilon^{\prime*}k)\left[\overline{u}\left(p^{\prime}\right)\slashed{\epsilon} \slashed{k}u\left(p\right)\right]\nonumber \\
 & +f_{4}(\epsilon k^{\prime})\left[\overline{u}\left(p'\right) \slashed{\epsilon}^{\prime*} \slashed{k}u\left(p\right)\right]+f_{5}\left[\overline{u}\left(p^{\prime}\right) \slashed{\epsilon} \slashed{\epsilon}^{\prime*}u\left(p\right)\right]+f_{6}(\epsilon\epsilon^{\prime*})\left[\overline{u}\left(p^{\prime}\right)u\left(p\right)\right]\nonumber\\
 & +f_{7}(\epsilon k^{\prime})(\epsilon^{\prime*}k)\left[\overline{u}\left(p^{\prime}\right)u\left(p\right)\right]+f_{8}(\epsilon\epsilon^{\prime*})\left[\overline{u}\left(p^{\prime}\right) \slashed{k}u\left(p\right)\right]+f_{9}\left[\overline{u}\left(p^{\prime}\right) \slashed{\epsilon} \slashed{\epsilon}^{\prime*} \slashed{k}u\left(p\right)\right].\label{eq:2}
\end{align}
In this case we get nine Compton structure functions $f_{i}$. Here,
all the dot products are defined in the four-dimensional space-time
with the following metric $\left(1,-1,-1,-1\right)$, and $u(p)$
denotes the Dirac spinor for free baryon. The choice of the basis
is not unique and can be defined differently (\cite{Babusci}), although
the evaluation of the polarizabilities based on the basis in Eq.(\ref{eq:1})
is more convenient. Here, the structure functions $R_{i}$ are directly
related to the electric, magnetic and spin-dependent polarizabilities
in the multi-pole expansion. This includes loops (up to the given order
of perturbation) and structure-dependent pole contributions, such
as tree-level baryon resonance excitations and Wess-Zumino-Witten
(WZW) (\cite{WZ}, \cite{W}) anomalous interaction. Keeping only the dipole-dipole and dipole-quadrupole transitions in the multipole expansion of the
Compton structure functions \cite{multi-1,multi-2,multi-3}, we connect the non-Born (NB) structure functions to the polarizabilities of the baryon in these simple equations: 
\begin{align}
{\displaystyle R_{1}^{NB}=} & \omega^{2}\alpha_{E1};\ \ R_{2}^{NB}=\omega^{2}\beta_{M1};\ \ R_{3}^{NB}=\omega^{3}(-\gamma_{E1E1}+\gamma_{E1M2});\nonumber \\
\label{eq:3}\\
{\displaystyle R_{4}^{NB}=} & \omega^{3}(-\gamma_{M1M1}+\gamma_{M1E2});\ \ R_{5}^{NB}=-\omega^{3}\gamma_{M1E2};\ \ R_{5}^{NB}=-\omega^{3}\gamma_{E1M2}.\nonumber 
\end{align}
Connecting Compton structure functions $f_i$ from Eq.\ref{eq:2} to $R_{i}$ from Eq.\ref{eq:1}, we have:
\begin{align*}
R_{1}= & {\displaystyle -\eta^{2}\chi m^{2}\left(f_{1}+f_{2}+f_{3}m(1+\eta+v)\left(1+\frac{\eta}{1+v}\right)\right)}-f_{5}m\left(1+v-\frac{2\eta^{2}\chi}{1+v}\right)-f_{6}m\left(1+v-\frac{\eta^{2}\chi}{1+v}\right)\\
 & +f_{7}\eta^{2}m^{3}\chi\left(1+v-\frac{\eta^{2}\chi}{1+v}\right)-f_{8}m^{2}\eta(1+v)\left(1+\frac{\eta(1+\chi)}{1+v}+\frac{\eta^{2}\chi}{(1+v)^{2}}\right)\\
 & -f_{9}\eta m^{2}\left(1+v+\eta+2\eta\chi\left(1+\frac{\eta}{1+v}\right)\right),\\
\\
R_{2}= & \eta^{2}m^{2}\left(f_{1}+f_{2}+f_{3}m(1+v+\eta)\left(1+\frac{\eta}{1+v}\right)\right)-f_{5}\frac{\eta^{2}m}{1+v}\\
 & -f_{7}\eta^{2}m^{3}\left(1+v-\frac{\eta^{2}\chi}{1+v}\right)+f_{9}\eta^{2}m^{2}\left(1+\frac{\eta}{1+v}\right),\\
\\
R_{3}= & 2\eta^{2}m^{2}\left(f_{1}-f_{3}m(1+v+\eta)-\chi\left(f_{2}+f_{3}m(1+v+\eta)\right)\right)+f_{5}m\left(1+v+\frac{2\eta^{2}\chi}{1+v}\right)+\frac{f_{6}\eta^{2}m\chi}{1+v}\\
 & -f_{7}\frac{\eta^{4}m^{3}\left(\chi^{2}-1\right)}{1+v}-f_{8}\eta^{2}m^{2}\chi\left(1+\frac{\eta}{1+v}\right)+f_{9}\eta m^{2}\left(1+v+\eta-2\eta\chi\left(1+\frac{\eta}{1+v}\right)\right),\\
\\
R_{4}= & \left(f_{5}+f_{6}\right)\frac{\eta^{2}m}{1+v}-\eta^{2}m^{2}\left(f_{8}+f_{9}\right)\left(1+\frac{\eta}{1+v}\right),\\
\\
R_{5}= & \eta^{2}m^{2}\left(f_{2}+f_{3}m(1+v+\eta)\right)-\left(f_{6}+2f_{5}\right)\frac{\eta^{2}m}{1+v}+f_{7}\frac{\eta^{4}m^{3}\chi}{1+v}+\eta^{2}m^{2}\left(f_{8}+2f_{9}\right)\left(1+\frac{\eta}{1+v}\right),\\
\\
R_{6}= & -\eta^{2}m^{2}\left(f_{1}-f_{3}m(1+v+\eta)\right)-f_{7}\frac{\eta^{4}m^{3}}{1+v},
\end{align*}
where $\chi=\cos\theta_{c.m.}$, $\theta_{c.m.}$ is a photon scattering
angle in the c.m.s. reference frame, $\eta=\frac{\omega}{m}$,
and $v=\frac{E}{m}$, where $E$ is the energy of baryon in c.m.s. As one can see from Fig.\ref{ff1}, the proton polarizabilities have almost no energy dependence below 50 MeV. 
The electric proton polarizability has strong, resonance-type dependence near the pion production threshold. Of course, we
need to add contribution from the resonances in the loops of Compton
scattering. Hence, we have borrowed the resonance loops results from
the small-scale expansion (SSE) approach (\cite{SSE}). If no $\Delta$-pole
contribution is added, the magnetic polarizability in Fig.\ref{ff1}
stays negative (diamagnetic) for almost all the energies. The $\Delta$-pole
contribution is large enough to shift $\beta_{p}(\omega)$ from negative
to positive (paramagnetic) values for energies up to 250 MeV. 
The pion loop calculations account for magnetic polarizability coming
from the virtual diamagnetic pion cloud, and the $\Delta$-pole resonance
contribution to $\beta_{p}(\omega)$ is driven by the strong paramagnetic
core of the nucleon. Thus, in relativistic ChPT up to one-loop order including the $\Delta$-pole and SSE contribution and extrapolated to zero energy, we have (in units of $10^{-4}$ $fm^3$):
\begin{align*}
\alpha_{p}= & (7.38\,(\pi-\mbox{loop})-0.95\,(\Delta-\mbox{pole})+4.2\,(\mbox{SSE}))=10.63;\\
\beta_{p}= & (-2.20\,(\pi-\mbox{loop})+3.0\,(\Delta-\mbox{pole})+0.7\,(\mbox{SSE}))=1.49.
\end{align*}

\begin{figure}
\begin{centering}
\includegraphics[scale=0.35]{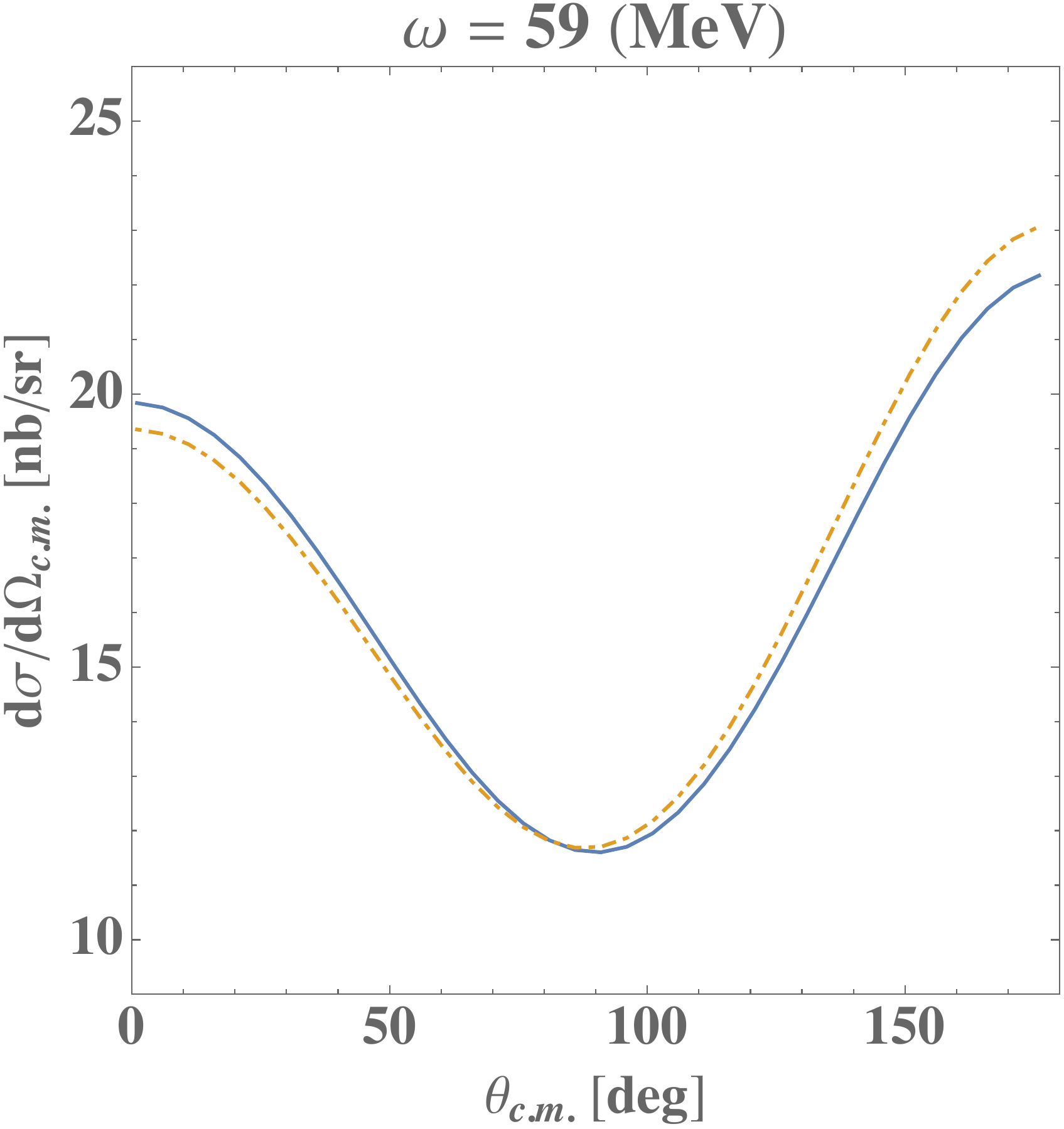} \includegraphics[scale=0.33]{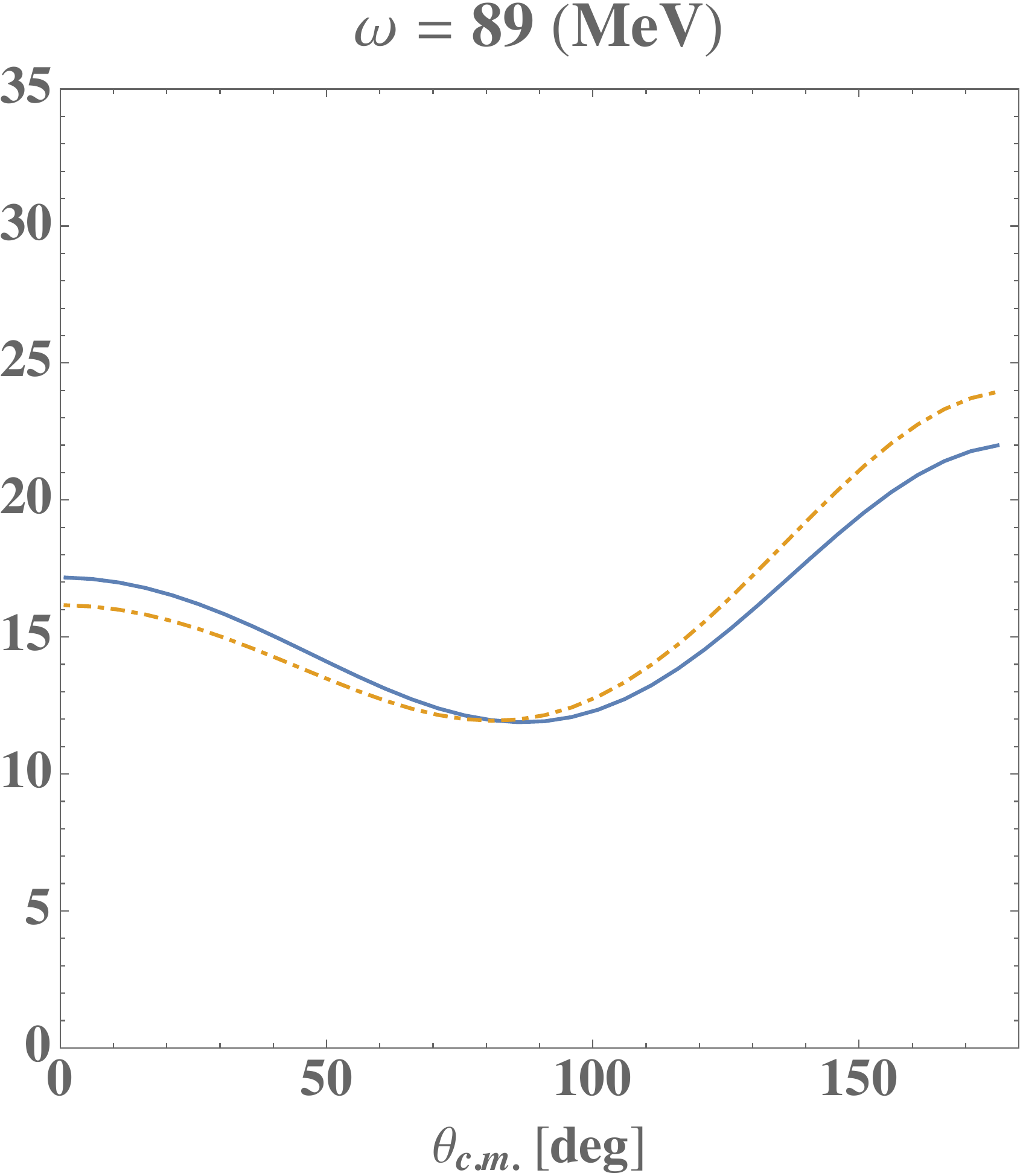}
\includegraphics[scale=0.33]{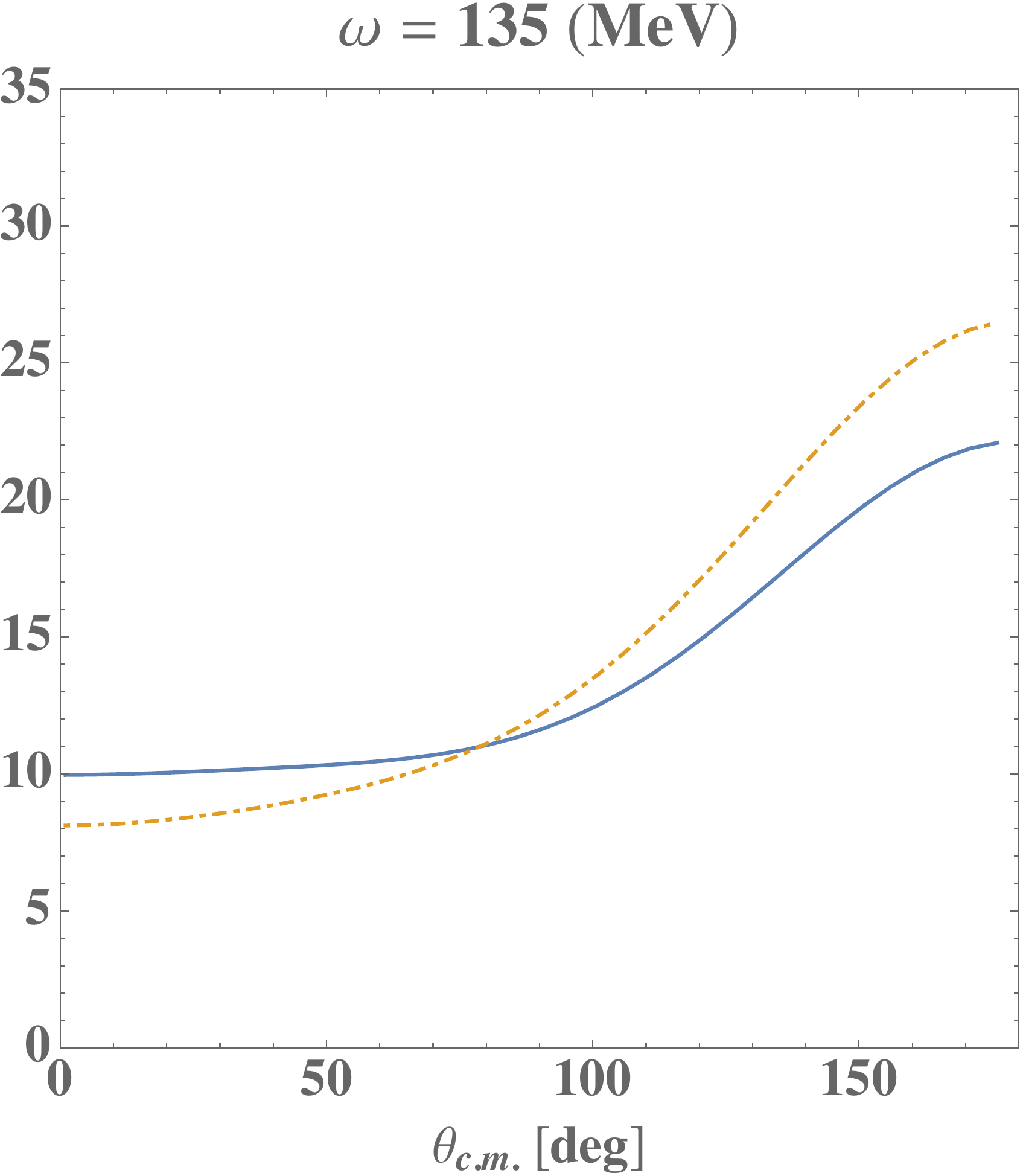}
\par\end{centering}
\caption{Unpolarized Compton scattering differential cross-sections computed
with CHM in c.m.s. for three photon energy $\omega$ values. The solid blue line corresponds to the
CHM predictions for Born with $\mathcal{O}\left(p^{3}\right)-\pi N$
loops and WZW-anomaly. The dot-dashed yellow line includes all the previous
contributions plus the $\Delta-$pole resonance. }

\label{fig2}
\end{figure}
Fig.\ref{fig2} clearly shows that $\Delta$-pole contribution becomes substantial with higher photon energy. We plan to address the inclusion
of $\Delta$-type resonances in the loops for polarizabilities and $\frac{d\sigma}{d\Omega}_{c.m.}$  in the future work.

\begin{acknowledgments}
This work is supported by the Natural Sciences and Engineering Research Council of Canada.
\end{acknowledgments}

\end{document}